\documentclass[12pt,preprint]{aastex}
\received{} \accepted{}
\journalid{000}{}
\begin{document}
\title{The Beylkin-Cramer Summation Rule and A New Fast Algorithm of Cosmic Statistics for Large Data Sets}

\author{Long-long Feng \footnote{fengll@pmo.ac.cn}}

\affil{Purple Mountain Observatory, Beijing West Road \#2, Nanjing
210008, China \\Joint Center for Particle Nuclear Physics and
Cosmology (JCNPC) \\Nanjing, 210093, China}

\begin{abstract}

Based on the Beylkin-Cramer summation rule, we introduce a new fast
algorithm that enable us to explore the high order statistics
efficiently in large data sets. Central to this technique is to make
decomposition both of fields and operators within the framework of
multi-resolution analysis (MRA), and realize theirs discrete
representations. Accordingly, a homogenous point process could be
equivalently described by a operation of a Toeplitz matrix on a
vector, which is accomplished by making use of fast Fourier
transformation.  The algorithm could be applied widely in the cosmic
statistics to tackle large data sets. Especially, we demonstrate
this novel technique using the spherical, cubic and cylinder counts
in cells respectively. The numerical test shows that the algorithm
produces an excellent agreement with the expected results. Moreover,
the algorithm introduces naturally a sharp-filter, which is capable
of suppressing shot noise in weak signals. In the numerical
procedures, the algorithm is somewhat similar to particle-mesh (PM)
methods in N-body simulations. As scaled with $O(N\log N)$, it is
significantly faster than the current particle-based methods, and
its computational cost does not relies on shape or size of sampling
cells. In addition, based on this technique, we propose further a
simple fast scheme to compute the second statistics for cosmic
density fields and justify it using simulation samples. Hopefully,
the technique developed here allows us to make a comprehensive study
of non-Guassianity of the cosmic fields in high precision cosmology.
A specific implementation of the algorithm is publicly available
upon request to the author.

\end{abstract}


\section{Introduction}

The data explosion in observational cosmology has been seen in past
decade. Though the rapid increasing both in quantity and quality
allow us to make a great improvement in measuring the global
properties of the universe and galaxy clustering, it poses new
challenges to extracting information from enormous data sets. For
instance, the early galaxy redshift survey published the catalog
only of a few thousand galaxies( QDOT; Lawrence et al. 1996,
Berkeley 1.2Jy; Fisher et al. 1995). So far, the number has been
raised to around 250,000 in the Two Degree Field Galaxy Redshift
Survey (2dFGRS) and $10^6$ in the Sloan digital sky survey £¨SDSS).
The upcoming galaxy redshift survey in a few years will reach to
$10^7$ by the Large Sky Area Multi-object Fibre Spectroscope
Telescope (LAMOST). Moreover, the state of arts for cosmological
numerical simulations are capable of producing galaxy clustering
models with $2160^3\simeq 1.0078 \times 10^{10}$ CDM particles in
the millennium simulation (Springer, et al., 2005), which have made
us in position to address some interesting astrophysical problems.
however, amenable to currently available numerical technique, it is
not always possible to make all of desirable statistical
measurements in those considerably large data.

The two-point correlation function and its Fourier counterpart, the
power spectrum are basic statistic measure to quantify the large
scale structure of the universe. The standard optimal estimator of
the correlation function (Landy \& Szalay 1993, Szapudi \& Szalay,
1998) requires $O(N^2)$ operations through counting pairs of
particles cycling over the whole sample, where $N$ is the number of
objects. For a large N, the CPU time in the pair-counting is a
paramount consideration in practice. In limited case, e.g. while
measuring the correlation function on small scales or reasonable
large bins, the pair-counting could be sped up to $O(N\log N)$ by
the double-tree algorithm (Moore et al. 2001). But this algorithm
degenerates to the naive $O(N^2)$ if all of scales are taken into
account. Complementary to the tree-based algorithm, Szapudi et al.
(2005) proposed a grid-based algorithm using the FFT technique. In
computation, it is scaled as $O(N_g\log N_g)$ on all the scales
though it would be ideal for large scales, here $N_g$ is the number
of grids.

The cosmic statistics higher than the second order characterizes the
Non-Gaussian features developed in the highly non-linear
gravitational clustering processes and provide a powerful probe to
gravitational collapse and merging of massive halos on small scales.
For instance, they could be applied to discriminate various
clustering models beyond the second order. And also, the encoded
information of biasing effects in the hierarchical clustering
scenario could be extracted from the higher order statistics to
place an extra constraints on the models. Recently, the rapid
increasing quantity and quality of both the 2dFGRS and SDSS have
stimulate great efforts in measuring high order statistics with
significantly improved accuracy. For instance, Jing \& Borner (2004)
measured the three-point correlation function in an early release
2dF100K sample, Wang et al (2004) devoted to the same subject to
test the conditional luminosity function model. Other approaches
included estimating the biasing factor from the bispectrum (Verde et
al, 2002) and moments of counts-in-cells in the 2dFGRS(Croton et al,
2004a,b).

However, the higher order statistics relies on large number of
parameters in the configuration space, which complicates both
analytical studies and practical measurements. Among those, the
counts-in-cells (CIC) and relevant conditional cumulants (Szapudi
2004, 2005) provide a relative simple non-Gaussian indicator.
Especially in the latter, it could be understood as degenerate
N-point correlation functions, or integrals of the monopole moment
of the bispectrum . While for the conventional CIC measurement using
cubic cells, to achieve reliable statistical significance, it
required a large number of random sampling and counting. To speed up
measurements, several fast methods which are optimal for different
purposes have been developed. In addition, an alternative approach
to cosmic statistics is using discrete wavelet transformation (DWT)
for cosmic fields, which enable us to probe the clustering features
both in the position and wavenumber spaces simultaneously. The DWT
method offers some sensitive statistical detectors to reveal the
non-Gaussian features of fields (Feng, Deng \& Fang, 2000, Feng \&
Fang 2000, Feng, Pando \& Fang 2001). For a comprehensive review of
the high order cosmological perturbation theory, we refer to
Bernardeau et al.(2002) and on the high order statistical methods,
refer to Szapudi (2005) .

This paper is to describe a new fast algorithm based on newly
developed technique in numerical mathematics (Beylkin \& Cramer,
2002). Actually, the algorithm is realized making using of a fast
summation rule within the framework of multi-resolution analysis
(MRA)(Beylkin \& Cramer, 2002), and could be applied widely to point
processes with the given filtered kernels. Furthermore, with the
MRA-CS scheme, we proposed a new method for calculations of the
second-order statistical measures such as the two-point correlation
function and variances of density fluctuations. We also noticed the
mostly recent paper by Thacker \& Couchmann (2005), where a fast
statistical measurement technique was proposed. Their approach is
similar to the technique described in our paper, however, they
adopted the 3rd order B-spline as an assignment function for mass
partition. We here present a more general method based on the
well-proven mathematics.

The paper is organized as follow, in \S 2, we give a general
formulation of the new fast algorithm. In particular, a new simple
algorithm for measuring the second order statistics is described in
\S 2.3. \S 3 presents numerical tests of the MRA-CS algorithm using
N-body simulations. Finally, we summary the paper and give
concluding remarks. In addition, as we make use of B-spline
biorthogonal bases in implementation of the algorithm, some useful
properties of B-splines are briefed in Appendix A.

\section{General Formulae}

\subsection{Filtered Density Fields}

The 3-dimensional distribution of galaxies can be modeled by a
spatial point process with density $n({\bf x})$ , which is formally
written as a sum of Dirac delta function $\delta_D^3({\bf x})$,
\begin{equation}
n({\bf x})=\sum_{i=0}^{N}w_i\delta_D^3({\bf x-x_i})
\end{equation}
where $N$ is the total number of samplinggalaxies, ${\bf x}_i$ is
the position of the $i$th particle and $w_i$ is its weight. For a
flux limited galaxy redshift survey, $w_i$ is given by the
luminosity selection function. For most of measurable statistical
quantities in the sample, we need to evaluate a discrete sum of
\begin{equation}\label{sum}
n_W({\bf x})=\sum_{i=1}^{N}w_i W({\bf x}-{\bf x}_i)
\end{equation}
where $W({\bf x})$ is a kernel. Alternatively, the above summation
can be written in terms of the original density distribution $n({\bf
x})$ convolved with the kernel $W({\bf x})$
\begin{equation}\label{densityw}
n_W({\bf x})=\int W({\bf x-x'})n({\bf x'})d^3{\bf x'}
\end{equation}
In the wavenumber space, the Fourier image reads
\begin{equation}
\hat{n}_W({\bf k})=\hat{W}({\bf k})\hat{n}({\bf k})
\end{equation}

Here we list some kernels frequently used in statistical analyses of
the cosmic density fields.

(1) Spherical top hat:
\begin{equation}
W_{sphere}(r,R)=\frac{1}{(4\pi/3)R^3}\theta(R-r)
\end{equation}
where $\theta(r)$ is the Heaviside step function, $R$ is the filter
radius. The Fourier counterpart is
\begin{equation}\label{sphk}
\hat{W}_{sphere}(k,R)=\frac{3}{k^3R^3}(\sin(kR)-kR\cos(kR))
\end{equation}
Unlike a sharp cutoff at the radius R in real space, top hat filter
in k-space is rather diffuse. This filter smooths over a finite
spherical volume of radius R around a point. Actually, it gives a
volume-averaged number density of particles, and thus counts the
total number of particles in a given spherical cell.

(2) Spherical shell top hat
\begin{equation}
W_{shell}(r,R)=\frac{1}{4\pi r^2}\delta_D(R-r)
\end{equation}
which defines the average surface density in the sphere shell at
distance $R$ from a given point. Its Fourier transformation gives
\begin{equation}
W_{shell}(k,r)=\frac{\sin(kr)}{kr}
\end{equation}
Obviously, it is related to the spherical top hat both in the real
and wavenumber space by
\begin{equation}
W_{shell}(\cdot, R)=\frac{1}{3R^2}\frac{d}{dR}(R^3W_{sphere}(\cdot,
R))
\end{equation}
For the statistical estimator of the two-point correlation function,
it is a basic quantity to be measured.

(3) Cubic top hat:
\begin{equation}
W_{cubic}({\bf x},{\bf L})
=\frac{1}{L_xL_yL_z}\theta(L_x/2-|x|)\theta(L_y/2-|y|)\theta(L_z/2-|z|)
\end{equation}
and in the $k$-space
\begin{equation}
\hat{W}_{cubic}({\bf k},{\bf L})=
\frac{\sin(k_xL_x)}{k_xL_x}\frac{\sin(k_yL_y)}{k_yL_y}\frac{\sin(k_zL_z)}{k_zL_z}
\end{equation}
where the vector ${\bf L}$ denoting for $\{L_x,L_y,L_z\}$ defines
the side length of cubic. Similar to the spherical top hat, the
cubic one gives a volume-averaged density in a given cubic.

(4) Cylinder top hat:

\begin{equation}
W_{cylinder}(\rho,z,R,h)=\frac{1}{\pi R^2
h}\theta(R-\rho)\theta(h/2-|z|)
\end{equation}
where $h$ is the height along axis of cylinder, $R$ is the radius of
circular cross section. In the $k$-space, we have
\begin{equation}
\hat{W}_{cylinder}(k_{\bot},k_z,R,h)=\frac{\sin(k_zh/2)}{k_zh/2}\int_0^1J_0(k_{\bot}R\sqrt{x})dx
\end{equation}
where $k_{\bot}=\sqrt{k_x^2+k_y^2}$

(5) Gaussian Filter:

The Gaussian filter behaves in the same way both in the coordinate
space and in the wavenumber space,
\begin{equation}
W_G(r,R)=\frac{1}{\sqrt{(2\pi)^3}R^3}\exp\bigl(-\frac{r^2}{2R^2}\bigr)
\end{equation}
\begin{equation}
\hat{W}_G(k,R)=\exp(-\frac{1}{2}k^2R^2)
\end{equation}

\subsection{The Beylkin-Cramer Fast Summation Rule in Multiresolution Analysis}

Central to multiresolution analysis (MRA) (Daubechies, 1992; Fang \&
Thews, 1998) is to express an arbitrary function at various levels
of the spatial resolutions, which forms a sequence of functional
spaces ${0}\subset \cdot\cdot\cdot \subset V_{-1} \subset V_0
\subset \cdot\cdot\cdot \subset L^2({\bf R})$ . Suppose a set of
functions $\{\phi(x-k)|k\in {\bf Z}\}$ forms an orthonormal basis
for $V_0$, dilated by a scale $2^{j}$ and translated by $2^{-j} k$
yields an orthogonal basis for $V_j$,
\begin{equation}
\{\phi_{j,k}(x)=2^{j/2}\phi(2^jx-k)\quad |k\in {\bf Z}\}
\end{equation}
where $\phi$ is called the basic scaling function. Projection of a
function $f\in L^2({\bf R})$ onto $V_j$ is approximation to $f$ at
the scale $j$, and converges to $f$ as increasing $j\rightarrow
\infty$. Without loss of generality, all of formulae hereafter are
written in one-dimensional forms. The generalization to
multi-dimensional space is straightforward.

In terms of scaling functions, we may make a decomposition of the
density distribution $n(x)=\sum w_i\delta(x-x_i)$ in the MRA at a
scale $j$.
\begin{equation}\label{denj}
n(x)=\sum_{l}s^j_l\phi_{j,l}(x)
\end{equation}
in which the scaling function coefficients (SFCs) $\{s^j_l\}$ are
given by the inner product
\begin{eqnarray}\label{sum}
s^j_l&=&\int n(x)\phi_{j,l}(x)dx\\ &=&
\sum_{i=1}^{N_p}w_i2^{j/2}\phi(2^jx_i-l)
\end{eqnarray}
which describes the density fluctuation filtered on the scale of
$1/2^j$ at position $l/2^j$ (Fang \& Thews, 1998, Feng \& Fang,
2004). Usually, the scaling functions is chosen to have a compact
support. In this case, the summation of cycling through the whole
sample reduces to sum up the contributions from neighbouring
particles around the position $l$.

Similarly, for a kernel $W(x,y)$, projection onto $V_j$ yields a
multiresolution representation $W \rightarrow W_j$, which has the
form (Beylkin \& Cramer,  2002),
\begin{equation}\label{wj}
W_j(x,y)=\sum_{l,m} w^j_{l,m}\phi_{j,l}(x)\phi_{j,m}(y)
\end{equation}
where
\begin{equation}\label{wjl}
w^j_{l,m}=\int W(x,y)\phi_{j,l}(x)\phi_{j,m}(y)dxdy
\end{equation}
Substituting eq.(\ref{denj}) and eq.(\ref{wj}) into
eq.(\ref{densityw}), we obtain the filtered density field at the
scale $j$ as
\begin{equation}\label{readnw}
n_{W}(x) \rightarrow n_{W}^j(x)=\sum_l \tilde{s}^{j}_{l}
\phi_{j,l}(x)
\end{equation}
with
\begin{equation}\label{ws}
\tilde{s}^{j}_{l} = \sum_{m}w^j_{l,m} s^j_m
\end{equation}
For a homogenous kernel $W$, $w^j_{l,m}=w^j_{l-m}$ is a Toeplitz
matrix, and conventionally its operation on the vector ${\bf s}^j =
\{s^j_m\}$ in eq.(\ref{ws}) could be accomplished using the fast
Fourier transformation technique (FFT).

As described above, we have a fast algorithm based on the
multiresolution analysis, which will be hereafter called MRA-CS
({\bf M}ulti-{\bf R}esolution {\bf A}nalysis for {\bf C}osmic {\bf
S}tatistics). Practically, once the orthonormal basis functions are
specified, the algorithm can be implemented in the following four
steps:

(1) computing the SFCs of the density field by the summation
eq.(\ref{sum});

(2) at a given scale $j$, computing the coefficients $w^j_{n}$ that
represent the kernel $W$ in eq.(\ref{wjl}), and then making discrete
Fourier transformation of $w^j_{n}$ to get the Green function
$\hat{w}^j_{n}$ in the wavenumber space;

(3) performing multiplications of the Toeplitz matrix $w^j_{l-m}$ to
the vector $\{s^j_m\}$ via the FFT;

(4) at any points, values of the filtered density field $n_W(x)$ can
be read out simply by the summation eq.(\ref{readnw}).

%Obviously, the above procedure is analogue to the Particle-Mesh
%gravity solver in N-body simulations, which is proceeding in
%following steps  (1) for a given particle distribution, distributing
%mass onto meshes using a mass assignment scheme (2) tabulating the
%Green function on meshes (3) solving the Possion equation using the
%FFT technique in wavenumber space and (4) making interpolation of
%force at particles.

\subsection{ The Fast Algorithm for Two-Point Correlation Function }

The two-point spatial correlation function $\xi({\bf
r})=<\delta({\bf x})\delta({\bf x+r})>$ of a density field describes
the lowest order statistical measure of departure of the matter
distribution from homogeneity. It is Fourier conjugate of the power
spectrum $P(k)=<|\delta({\bf k})|^2>$,
\begin{equation}\label{2pcfps}
{\xi}(r)=\frac{1}{2\pi^2}\int_0^{\infty}\hat{W}_{shell}(k,r)P(k)k^2dk
\end{equation}
In practical ststistical analysis, there have been several
edge-corrected estimators of the two-point correlation function
proposed so far. The widely used one is that given by Landy and
Szalay, which takes the form
\begin{equation}\label{LS}
\hat{\xi}_{LS}(r)=\frac{DD-2DR+RR}{RR}
\end{equation}
where $DD$ is the count of pairs in the catalog within the interval
$[r-0.5dr,r+0.5dr]$, $RR$ is the number of pairs in a random sample,
and $DR$ the number of pairs between the catalog and the binomial
random sample in the same interval.

Based on the MRA-CS technique, we describe here a fast and simple
algorithm to estimate the number of pairs. Firstly, we consider
counting the $DD$-pair in the catalog. According to
eq.(\ref{readnw}), the pair count in the shell of $(r,r+dr)$ from a
particle $i$ is approximated by
\begin{equation}
n_{W}^j(x_i)= dV \sum_l \tilde{s}^{j}_{l}\phi_{j,l}(x_i)
\end{equation}
where $dV=4\pi r^2dr$ is the differential volume element of
spherical shell if $dr$ is small enough. Since $dV$ is a common
factor for the pair counts $DD$, $DR$ and $RR$, it will be dropped
hereafter. Taking the sum over the whole sample, the total number of
pairs is
\begin{equation}
DD=\sum_{i=1}^{N}w_i\sum_l \tilde{s}^{j}_{l} \phi_{j,l}(x_i)
\end{equation}
which can be written in an equivalent form,
\begin{equation}
DD=\sum_l\tilde{s}^{j}_{l}\int dx  \phi_{j,l}(x)n(x)
\end{equation}
where $n(x)$ is the number density given by eq.(1). Using the
decomposition eq.(\ref{denj}) and orthogonality of the base
functions $\{\phi^{j}_{l}\}$, we have
\begin{equation}
DD=\sum_l \tilde{s}^{j}_{l} s^{j}_{l}={\bf \tilde{s}}^j\cdot{\bf
s}^j
\end{equation}
The above equation means that counting the number of pairs can be
simplified to a vector product of ${\bf s}^j=\{s^j_l\}$ and ${\bf
\tilde{s}}^j=\{\tilde s^j_l\}$. The similar algorithm can be applied
for the $DR$ and $RR$ pairs. For instance, in counting the $DR$
pairs, we may still use eq.(29), but $\{s^j_l\}$ are given instead
by the decomposition coefficients calculated for the given random
sample.

We emphasis here that, in eqs.(25)-(28), $\tilde{s}^{j}_{l}$ is
obtained from a convolution of ${s}^{j}_{l}$ with the sphereical
shell top hat $W_{shell}(k,r)$. Actually, this algorithm allows us
to calculate the correlation function without counting particles in
a series of spherical shells around each particle. Namely, it omits
binning in the radial direction of spheres, and thus is bin width
independent.

On the other hand, if we use the sphereical top hat $W_{sphere}$,
the pair-counting is performed in the sphere of radius $r$ centered
on each particle. In this case, the correlation function estimated
from eq.(\ref{LS}) is just what we refer as the integral or
volume-averaged two-point correlation function, which is related to
$\xi(r)$ by
\begin{equation}
\bar{\xi}(r)=\frac{3}{r^3}\int_0^r\xi(x)x^2dx
\end{equation}
Analogue to eq.(\ref{2pcfps}), we have
\begin{equation}
\bar{\xi}(r)=\frac{1}{2\pi^2}\int_0^{\infty}\hat{W}_{sphere}(k,r)P(k)k^2dk
\end{equation}
where $\hat{W}_{sphere}(k,r)$ is the spherical top hat window
function given by (eq.(\ref{sphk})).

For the second order variance of filtered density fluctuations
\begin{equation}
\sigma^2(\cdot) = <\delta_W^2(\cdot)> =\frac{1}{(2\pi)^3}\int
|W_{filter}({\bf k},\cdot)|^2P(k)d^3{\bf k}
\end{equation}
where the dot $\cdot$ denotes for a set of parameters specifying the
spatial shape of the window function. Similar to the above
derivation, we may show that it can be worked out simply by
\begin{equation}
\sigma^2(\cdot) = \sum_l \tilde{s}^{j}_{l} \tilde {s}^{j}_{l}=|{\bf
\tilde{s}}^j|^2 -1
\end{equation}

Obviously, in the spirit of this method, it is possible to
generalize the MRA-CS algorithm for higher order statistics. In that
case, we need to deal with integrals like $\Gamma^{j}_{lmn}=\int
\phi_{j,l}\phi_{j,m}\phi_{j,n} dx$, which are currently referred as
the connection coefficients. For the detail discussions, we will
give in the forthcoming paper.

\section{Numerical Tests}

In the last section, we have described a fast algorithm MRA-CS for
filtered cosmic fields. The direct application is the
counts-in-cells with arbitrary geometries and the second order
statistical measures such as the two-point correlation funcction and
the variance of filtered density fields. To justify to what extent
this numerical scheme works well, we make numerical tests using
N-body simulation samples. We analyzed the $z=0$ output of N-body
simulations by the Virgo Consortium (Kauffman et al. 1999) in a LCDM
model specified by the cosmological parameters $\Omega_m=0.3$,
$\Omega_{\Lambda}=0.7$, $\Gamma=0.21$, $h=0.7$, $\sigma_8=0.85$ and
force soft length $20h^{-1}$kpc. The side length of the simulation
box was $239.5$h$^{-1}$Mpc, and the number of CDM particles was
$256^3$.

Analytical form of the MRA-CS scheme presented in the last section
is based on the multi-resolution analysis using compactly supported,
orthonormal bases involving a single scaling function. In practical
applications, we can alternatively adopt a biorthogonal system
including a pair of dual scaling functions. In this paper, we choose
the central B-splines in the implementation. As the B-splines do not
form an orthogonal basis, they may be parts of a biorthogonal
system. The corresponding dual scaling function could be constructed
as that spanned by the B-splines, and is not compactly supported. In
this case, some changes need to be made in the algorithm. For this
purpose, we present some useful properties and formulae of the
B-splines in Appendix A.

The naive measurement of counts in cells can be made by direct
counting. To test the MRA-CS algorithm, a direct method is to
compare results obtained by the MRA-CS with the exact counting. In
what follow, we perform numerical tests for the spherical, cubic and
cyclinder counts in cells, respectively. In the calculations, we
applied the 5th order B-splines. Meanwhile, alternative to the
method described in Appendix.2, according to eqn.(A10), we adopted
the Green function simply by $\hat{W}(2\pi{\bf \xi})\cdot
|\hat{\gamma}|^2=\hat{W}(2\pi{\bf \xi})(\hat{\beta}(\xi)/a(\xi))^2$
tabulating on a discrete grid, where $\hat{\beta}(\xi)$ and $a(\xi)$
are given by eqs.(A5) and (A8) respectively.

In the first numerical experiment, we randomly placed a sphere
center in a poisson sample with $256^3$ particles in a $256^3$ grid,
and then measure the numbers of particles within a sequence of the
concentric spherical shells at different radius. Figure 1 plots the
volume-average density vs. radius for three randomly placed
concentric sphere shells. Clearly, the results calculated from our
fast algorithm shows an excellent agreement with the exact values.

We perform further numerical experiment to test the counts-in-cells
with different geometries using the Virgo simulation sample in a
$512^3$ grid. We have done two sets of the counts-in-cells tests.
One was made for fixing the center of cells in a selected particle
within a massive halo, and another is for a low density region. The
results for the counts in cubic cells are demonstrated in figure 2,
where each panel is for cubes with a specific value of the aspect
ratio. The horizonal axis is denoted by the equivalent radius of
spheres with the same volumes as the cubic cells. Obviously, except
for those at scales of a few grids, the MRA-CS algorithm matches
with the exact counting very well. Similar results are displayed in
figure 3 for the spherical and cylinder counts-in-cells.

It is worthwhile to note that, at small scales, there are visible
differences between the MRA-CS and exact counting. Actually, in the
MRA-CS scheme, the Fourier transformation of the B-spline of order
$n$ is a low pass filter with a sharp suppression $1/k^{n+1}$ of
power. Consequently, the Poisson shot noise and fluctuations at
short wavelengths will be suppressed. This kind of behavior can be
viewed clearly in both figure 2 and figure 3. In case for the
counts-in-cells within the massive halo (the upper curve in each
panel), as the effect of shot-noise is not significant at small
scales, such a suppression leads to the volume-average densities
calculated from the MRA-CS scheme being systematically lower than
the direct counting as indicated in figure 2. While in the less
dense region(the lower curve in each panel), the shot noise
dominates at small scales, there are significant fluctuations in the
volume-averaged density obtained by direct counting. In comparison,
the MRA-CS gives slightly smooth variations of the density with
radius.

% Similar results are demonstrated in fig.(2), where we made the
% measurements in five sets of randomly throwing concentric spherical
% shells. To have a good view of errors, the volume-average densities
% are plotted in logarithmic scales, Actually, on scales larger than
% 10 grids, the accuracy of measurements are mostly better than
% $1.0\%$, while down to 5-10 grids, are no less than $5.0\%$.

To have a visual inspection of accuracy of the MRA-CS method, we
measure the counts-in-cells using $256^3$ spherical cells randomly
thrown in the simulation sample, and make a scatter plot in figure 4
to compare the volume-averaged density calculated from the MRA-CS
versus those by the direct counting. In order for a good view of
errors, the volume-averaged densities are plotted in logarithmic
scales. The upper and lower panel display the results for the cell
sizes of $R=5$h$^{-1}$Mpc and $R=10$h$^{-1}$Mpc, respectively.
Figure 4 shows more clearly the effects of shot noise occurred in
both cases. In the case $R=5$h$^{-1}$Mpc, the larger scatters appear
in low density regions where sparse sampling makes shot noise more
significant than in dense regions. Increasing the size of cell will
lead to, on average, more sampling particles in cells, and
consequently, the overall scatters tend to be squashed rather
smaller. In other words, for a given density, large cells contain
more particles than those in small cells, and the effect of Poisson
noise in the large cells should be less significant than in the
small cells accordingly. This kind of phenomena has been actually
illustrated in figure 4 by comparing the case of $R=5$h$^{-1}$Mpc
(upper panel) with $R=10$h$^{-1}$Mpc (lower panel).

Moreover, in figure 5, we display the probability distribution
functions (PDF) of the counts in cells for $R=5$h$^{-1}$Mpc and
$R=10$h$^{-1}$Mpc with three different shapes. In calculations, we
used 180 bins to make the PDFs.  Basically, there is an excellent
agreement of the MRA-CS calculation and the direct counting. In the
case $R=5$h$^{-1}$Mpc, the PDFs compiled from the direct counting
shows slight fluctuations in the low density regions $\rho_V<0.5$
due to the effect of shot noise. In comparison, such fluctuations
have been smoothed out in the MRA-CS calculations. Whereas in case
of $R=10$h$^{-1}$Mpc, the direct counting and MRA-CS shows a very
good agreement as shown in figure 5. It indicates again that the
shot noise could be reduced with increasing cell sizes. This
demonstration does show how a reliable statistics could be worked
out through the MRA-CS technique.

In further applications of the MRA-CS scheme, we measured the second
order statistics in the Virgo sample using the fast algorithm
described in \S 2.3. A $512^3$ grid has been used, and the physical
grid size is thus $0.47$h$^{-1}$Mpc. The results include the spatial
two-point correlation function in figure 6, the volume-averaged
two-point correlation function in figure 7, the variance of density
fluctuations by the top-hat filter in figure 8 and the Gaussian
filter in figure 9, respectively. In all figures, we also plot the
theoretical curves using the fitting formulae of nonlinear power
spectrum in LCDM models (Smith et al, 2003). The agreement of our
fast algorithm with the theoretical model looks quite perfect in
scales down to the grid size.

The current implementation of the MRA-CS algorithm is written in
Fortran-90. Compiling on a SGI Altix350 workstation of 16
Intel$\circledR$ Itanium 2 1.5G CPUs with the OPENMP
parallelization, it runs at the speed of 1.6 seconds per $256^3$
counts-in-cells using B-splines of degree $n=5$. By the scaling, a
massive counts-in-cells measurement with $10^9$ sampling cells could
be done with 2 minutes in the same machine. In the computations, the
Intel$\circledR$ Math Kernel Library (MKL) has been also applied to
perform the fast Fourier transformations.

Finally, it is necessary to point out that, the central idea in the
MRA-CS scheme is to find a finite and smooth representation of
discrete particle distribution, which is a singular function as
given by the Dirac delta function in eqn.(1). In principle, the
higher is order of B-splines,  the better is approximation to the
Dirac function (see Appendix 3). However, as computational cost is
scaled as $(n+1)^d$ where $n$ is the order of B-splines and $d$ is
the dimension of space, the higher order B-splines may leads to less
efficient in the algorithm. To set up a compromise between them, we
have numerically tested the dependence of accuracy on orders of
B-splines. The result indicates that for $n \ge 5$, there are no
visual differences at scales larger than the grid size. Accordingly,
we have used the fifth order of B-splines throughout this paper.

\section{Discussions and Concluding Remarks}

We have described a fast algorithm of cosmic statistics based on
multiresolution analysis, in which fields or operators are
decomposed in term of a set of orthonormal compactly supported bases
at various levels of details. This method naturally leads to a fast
summation rule, and could find wide applications in different areas.
In application to cosmic statistics, we made an implementation of
the MRA-CS scheme using high order B-spline biorthogonal bases and
justified its efficiency and reliability by the numerical
experiments. Using spherical,cubic and cylinder counts in cells, we
found that the MRA technique yields good estimations of the
counts-in-cells. On small scales, there appear some discrepancies in
comparison with direct counting. In statistics, it is not negative.
Actually, weak signals in a point process suffer from the Poisson
shot noise. However, in the MRA-CS scheme, those high frequency shot
noises could be suppressed by the virtue of the fast decaying
behavior of sharp low pass filters.

Based on the MRA-CS scheme, we proposed further a fast algorithm to
estimate the two-point spatial correlation function. Strikingly, we
proved that, taking average of pair counts over a sample can be
simplified to a vector product. That is, the averaged pair counts
can be obtained without counting. Is is also shown that this method
can be generalized to calculate other second order statistical
quantities such as the variances of cosmic fields filtered with
arbitrary window functions. Using the simulation sample, we
demonstrate an excellent agreement between our MRA-CS method and the
theoretical expectation.

The MRA-CS algorithm has several advantages,

(1) It is a grid-based algorithm making use of the FFT technique,
and thus is significantly faster than the conventional counts in
cells methods.

(2) In particular, the algorithm does not slow down while extending
to large scales. Actually, the number of operations required using
this method is independent of cell sizes. Therefore, this method
enable us to perform a massive sampling to reduce statistical
errors.

(3) The algorithm can be, in principle, applied to sampling cells
with arbitrary shapes, so as their spatial configurations could be
well defined mathematically.

When measuring the counts-in-cells in real galaxy samples, there are
two observational effects needed to be taken into account. The first
is volume incompleteness due to bright star masks and complicated
survey geometry. In the former, holes around bright stars should be
excavated, and in the latter some sampling cells beyond boundary
contain some fraction of space that are not in the survey volume.
The another effect is spectroscopic incompleteness. There are
several schemes suggested so far to correct galaxy counts in cells
for real samples ( e.g. Efstathiou et al. 1990, Croton et al. 2004).
For instance, in the Croton et al.'s approach applying to the 2dFGRS
catalogue, they first estimated the combined spectroscopic and
volume completeness factor $f$ using the survey mask and then
compensate the missed volume by expanding the cell such that the
resized incomplete cell has the same volume as the ideal fully
complete cell. Subjecting to the numerical tests, this method is
likely to give reasonable results.

This paper makes no attempt to discuss the issue of incompleteness
in detail. Here, we only brief a simple scheme to tackle this
problem. Practically, we may generate a uniform particle
distribution with the survey mask applied, and then compute its
SFCs. When estimating galaxy count within a cell in the real data
using the MRA-CS method, the same measurement is also performed in
the marked uniform sample. Actually, measuring the counts-in-cells
in the marked uniform sample may yield a good estimation of the
completeness factor $f$ so as it is dense enough. Explicitly, the
completeness factor $f$ for a cell is equal to its volume-averaged
density if the mean density of the unmarked uniform sample is taken
to be unity. It is basically an alternative to the current
Monto-Carlo method for computing irregular spatial volumes. In the
proceeding step, we may fix an acceptable minimum value of the
completeness factor $f_c$, sampling cells with $f< f_c$ are
excluded, otherwise, we may apply Croton et al's volume compensation
algorithm for galaxy counts in the re-scaled cells.

Statistically, the more information could be extracted by sampling
in larger configuration spaces of counts-in-cells. In fact, the
non-Gaussianities characterized by the high order statistics in the
cosmic fields are produced by the highly non-linear mode-mode
coupling, which may manifest itself in gravitational collapse or
merging of dark matter halos. Therefore, examining the shape
dependence of high order statistics might be helpful to understand
gravitational clustering pattern, e.g., the topology and large scale
structure, morphologies of voids, filaments and their networks, and
the origin of biasing. It is expected that the MRA-CS technique
developed in this paper could provide an efficient tool for
investigating the higher order statistical features in the large
scale structure of the universe.

The MRA-CS scheme has been implemented in a set of subroutines,
which is publicly available upon request to the author.

\acknowledgments  LLF thanks Pan Jun, Cai Yanchuan for helpful
discussions and comments, and also acknowledges the Institute for
Pure and Applied Mathematics in the University of California, Los
Angeles for hospitality during the program "Grand Challenge Problems
in Computational Astrophysics" (Mar.15 - May.25 2005) where this
work was initiated. This work was supported from the National
Science Foundation of China through grant NSFC 10373012. The
simulation in this paper were carried out by the Virgo
Supercomputering Consortium using computers based at Computering
Center of the Max-Planck Society in Garching and at the Edinburgh
Parallel Computing Center. The data are publicly available at
http://www.mpa-garching.mpg.de/Virgo.

\appendix

\section{B-splines }

\subsection{Definition and Basic Properties}

B-splines are symmetrical, bell shaped functions constructed from
the $n+1$ fold convolution of a rectangular pulse $\beta^{(0)}(x)$:
\begin{equation}
\beta^{(0)}(x)=\left\{ \begin{array} {r@{\quad \quad}l} 1 & if
|x|<0.5 \\0.5 & if |x|=0.5 \\ 0 & otherwise \end{array} \right.
\end{equation}
\begin{equation}
\beta^{(n)}(x)=(\beta^{(0)}_{+1}\circ
\beta^{(0)}_{+2}\circ\cdot\cdot\cdot
\beta^{(0)}_{+n}\circ\beta^{(0)}_{+(n+1)})(x)
\end{equation}
where the symbol $\circ$ denotes for operation of convolution.
$\beta^{(n)}(x)$ is a piecewise polynomial of degree n,
\begin{equation}
\beta^{(n)}(x)=\sum_{l=0}^{n+1}(-1)^l{n+1 \choose
l}\frac{(x+\frac{n+1}{2}-l)^n_{+}}{n!}
\end{equation}
In practical calculation, the B-splines could be obtained using
recursion over the spline order,
\begin{equation}
\beta^{(n)}(x)=\frac{(n+1)/2+x}{n}\beta^{(n-1)}(x+1/2)+\frac{(n+1)/2-x}{n}\beta^{(n-1)}(x-1/2)
\end{equation}

The Fourier transformation of $\beta^{(n)}(x)$ is related to the
$n+1$ fold convolution construction of the B-splines,
\begin{equation}
\hat\beta^{(n)}(\xi)=\Bigl(\frac{\sin \pi\xi}{\pi\xi}\Bigr)^{n+1}
\end{equation}

The $n$-th order B-spline has a compact support of $[-\frac{n+1}{2},
\frac{n+1}{2}]$. As $\beta$ is not orthogonal to its translates, we
apply the orthgonalizatio scheme to construct its dual base
$\gamma$, such that
\begin{equation}\label{orth}
\int_{-\infty}^{\infty}\beta^{(n)}(x-k)\gamma^{(n)}(x-l)dx=\delta_{kl}
\end{equation}
It follows from above orthogonality condition that, in the
wavenumber space,
\begin{equation}\label{gammann}
\hat{\gamma}^{(n)}(\xi)=\frac{\hat\beta^{(n)}(\xi)}{a^{(n)}(\xi)}
\end{equation}
with the periodic function
\begin{equation}
a^{(n)}(\xi)=\sum_{l=-\infty}^{\infty}|\hat{\beta}^{(n)}(\xi+l)|^2=\sum_{l=-n}^{n}\beta^{(2n+1)}(l)e^{i2\pi
l\xi}
\end{equation}
It is noted that the dual scaling function $\gamma$ is not compactly
supported.

\subsection{Kernel Representation Using B-spline Biorthogonal Bases}

For a given kernel $W(x,y)$, it could be decomposed in terms of
B-splines $\beta$ (Beylkin \& Cramer, 2002),
\begin{equation}
W(x,y)=\sum_k\sum_l w_{k-l}\beta(x-k)\beta(y-l)
\end{equation}
it follows from the orthogonality eq.(\ref{orth}) that the
coefficients $\{w_n\}$ can be found formally by
\begin{equation} \label{coff-gamma}
w_n=\int \gamma(x-n)(W\circ\gamma)(x)dx
\end{equation}
Since $\gamma$ can be expressed as a linear combination of
B-splines, we have
\begin{equation}\label{linear}
\gamma(x)=\sum_{-\infty}^{\infty}q_k\beta(x-k)
\end{equation}
Substituting eq.(\ref{linear}) into eq.(\ref{coff-gamma}), we have
\begin{equation}\label{w-zeta}
w_n=\sum_m \tilde{q}_{n-m}\zeta_m
\end{equation}
in which,
\begin{equation}\label{qq}
\tilde{q}_m=\sum_k q_kq_{m+k}
\end{equation}
and
\begin{equation}\label{zeta}
\zeta_n=\int \beta(x-n)(W\circ\beta)(x)dx
\end{equation}

Taking Fourier transformation on both sides of eq.(\ref{linear}), we
may see that the coefficients $\{q_n\}$ are Fourier coefficients of
the function $1/a(\xi)$, and the coefficients $\{\tilde{q}_n\}$ in
eq.(\ref{qq}) are Fourier coefficients of the function
$1/[a(\xi)]^2$. Define the trigonometric series $\hat{w}$ and
$\hat{\zeta}$ by
\begin{equation}\label{fft-w}
\hat{w}(\xi)=\sum_{-\infty}^{\infty}w_ne^{i2\pi n\xi}
\end{equation}
and
\begin{equation}\label{fft-zeta}
\hat{\zeta}(\xi)=\sum_{-\infty}^{\infty}\zeta_ne^{i2\pi n\xi}
\end{equation}
Thus, Fourier transformation of eq.(\ref{w-zeta}) implies
\begin{equation}\label{fft-wzeta}
\hat{w}(\xi)=\frac{\hat\zeta(\xi)}{[a(\xi)]^2}
\end{equation}
Obviously, eqs.(\ref{fft-w})(\ref{fft-zeta}) and (\ref{fft-wzeta})
formulate an algorithm of obtaining $w_n$ directly from $\zeta_n$.

Let the central B-spline $\beta$ is of odd degree $M-1$, then
support of $\beta(x)$ is $\{x:|x|\leq M/2\}$. Hence, eq(\ref{zeta})
can be written in form of
\begin{equation}\label{int-zeta}
\zeta_n=\int^{M}_{-M}B(x)W(x+n)dx
\end{equation}
where $B(x)=\int \beta(x+y)\beta(y)dy$ is the autocorrelation of
$\beta(x)$, which is supported on the interval $\{x:|x|\leq M\}$. If
the kernel $W(x,y)$ does not have singularity in whole axis $\bf R$,
just as either spherical or cubic kernel considered in this paper,
the integral eq.(\ref{coff-gamma}) is convergent, and only finite
number of the coefficients $\zeta_n$ are nonzero. For the singular
operator such as that of the power law $1/x^{\alpha}$, the
regularization procedure is necessary, the integral in
eq.(\ref{coff-gamma}) may fail to converge for all integer $n$,
because $\gamma(x-n)$ fails to vanish in a neighborhood of the
singularity of the kernel $W$. The detail for regularization scheme
based on the multiresolution approach could be found in Beylkin and
Cramer (2002).

\subsection{Approximation to the $\delta$-Function}

Let $n(x)=\delta(x-x_0)$ in eq.(\ref{denj}), and the scaling
function is taken to be B-splines,
$\phi_{j,l}(x)=\beta^{(n)}_{j,l}(x)$,  using the dual of $\beta$,
then we have the finite representation of $\delta$-function from
eq.(\ref{sum})
\begin{equation}
\delta(x-x_0) \rightarrow n^{j}_0(x)=\sum_{k\in {\bf
Z}}\beta^{(n)}_{j,k}(x_0)\gamma^{(n)}_{j,k}(x)
\end{equation}
It could be proven that, if $w$ is a test function, $w \in
C^{\infty}$ , then
\begin{equation}
w(x_0)=\int_{-\infty}^{\infty} n^{j}_0(x)w(x)dx + O(h^{n+1})
\end{equation}
where $h=2^{-j}, j>0$ (Beylkin, 1995).

\newpage

%fig1
\begin{figure}
%\begin{center}
%\vspace{16.0cm}
%\special{psfile=f1.ps  hoffset=-0  voffset=-10 hscale=70 vscale=70}
\centering
\includegraphics[width=14.0cm,angle=0]{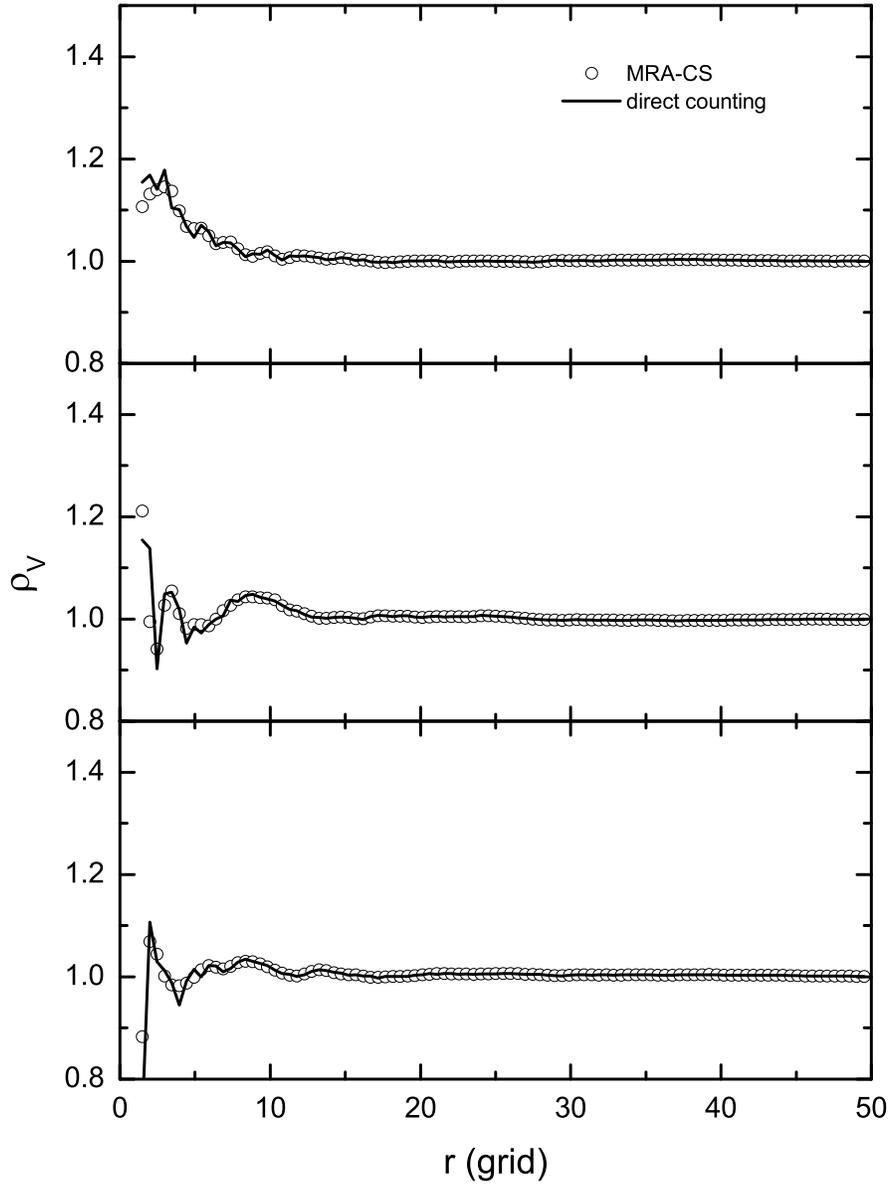}\vspace{-1.0cm}
\caption {Spherical counts in cells:  volume-averaged density varies
with radius measured in a sequence of concentric spherical shells
placed in a random sample with $256^3$ particles.  The MRA-CS
results(circules) are compared with the direct counting (solid
line). The upper, central and lower panels display the results in
three numerical experiments respectively}
%\end{center}
\end{figure}

%fig2
\begin{figure}
%\begin{center}
%\vspace{16.0cm}
%\special{psfile=f1.ps  hoffset=-0  voffset=-10 hscale=70 vscale=70}
\centering
\includegraphics[width=18.0cm,angle=0]{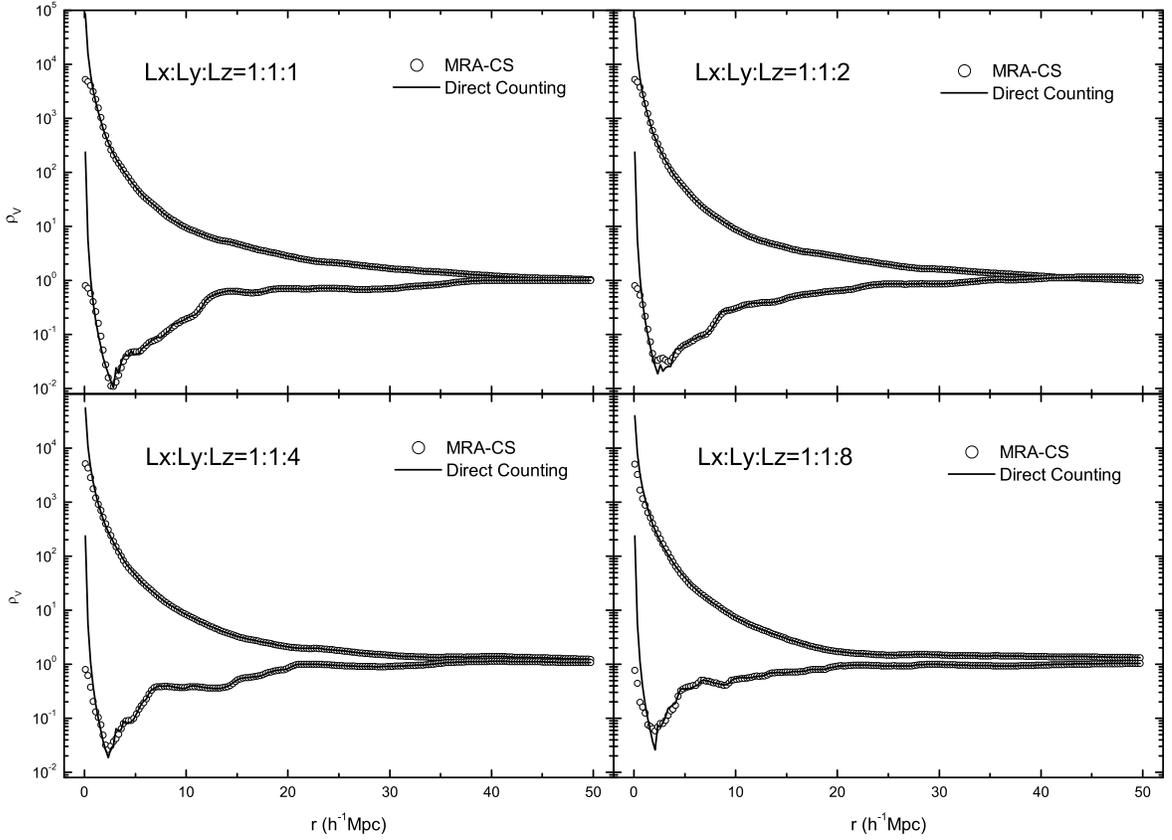}\vspace{-1.0cm}
\caption { The volume-averaged density measured in the Virgo
simulation sample as a function of cell size for a fixed center of
cells. Each panel corresponds to a given aspect ratio as indicated
in the figure, and plots two sets of the counts in cells
measurements. One is made in a massive dark matter halo (upper
curves), and another is for a low density region (lower curves). The
figure compares the results obtained from the MRA-CS algorithm (open
circules) with the direct counting (solid line).}
%\end{center}
\end{figure}

%fig3
\begin{figure}
%\begin{center}
%\vspace{16.0cm}
%\special{psfile=f1.ps  hoffset=-0  voffset=-10 hscale=70 vscale=70}
\centering
\includegraphics[width=18.0cm,angle=0]{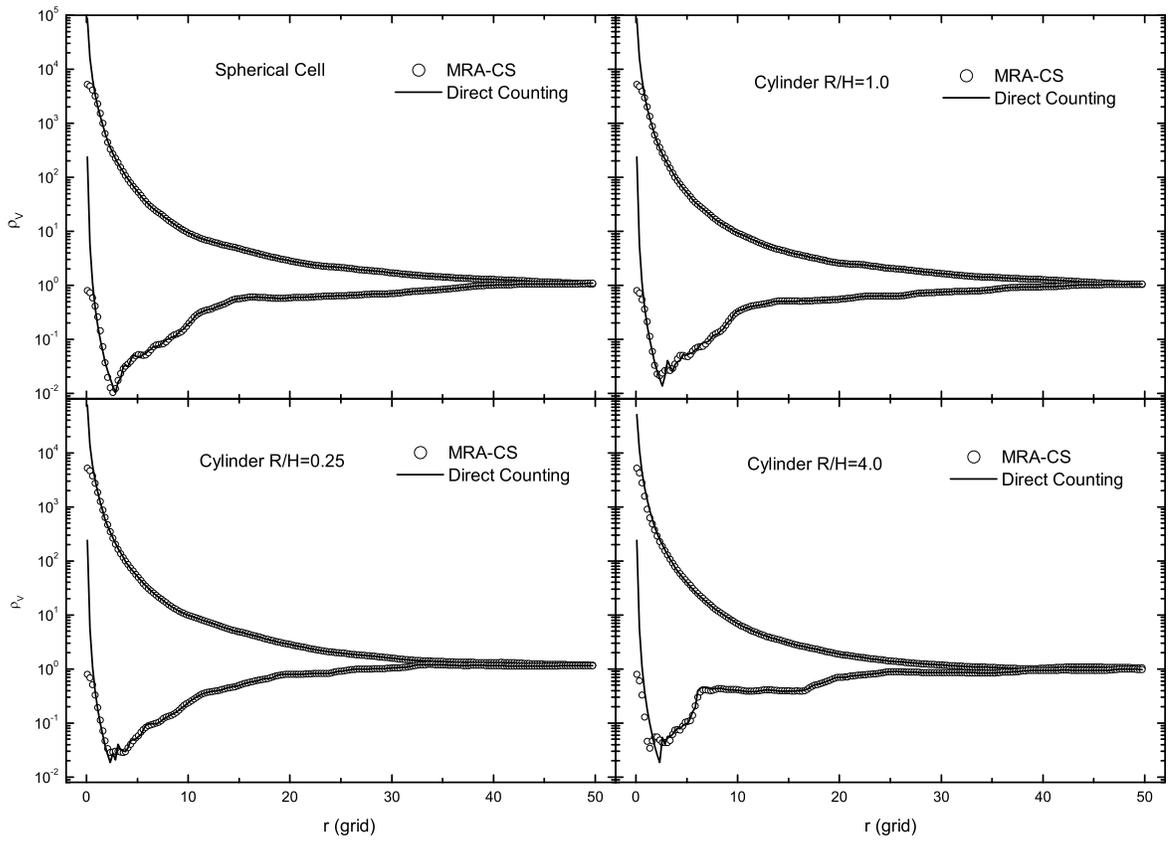}\vspace{-1.0cm}
\caption {Same as fig.(2), but for spherical and cylinder counts in
cells. For the cylinder cells, the aspect ratio is indicated in each
panel.}
%\end{center}
\end{figure}

%fig4
\begin{figure}
%\begin{center}
%\vspace{16.0cm}
%\special{psfile=f1.ps  hoffset=-0  voffset=-10 hscale=70 vscale=70}
\centering
\includegraphics[width=15.0cm,angle=0]{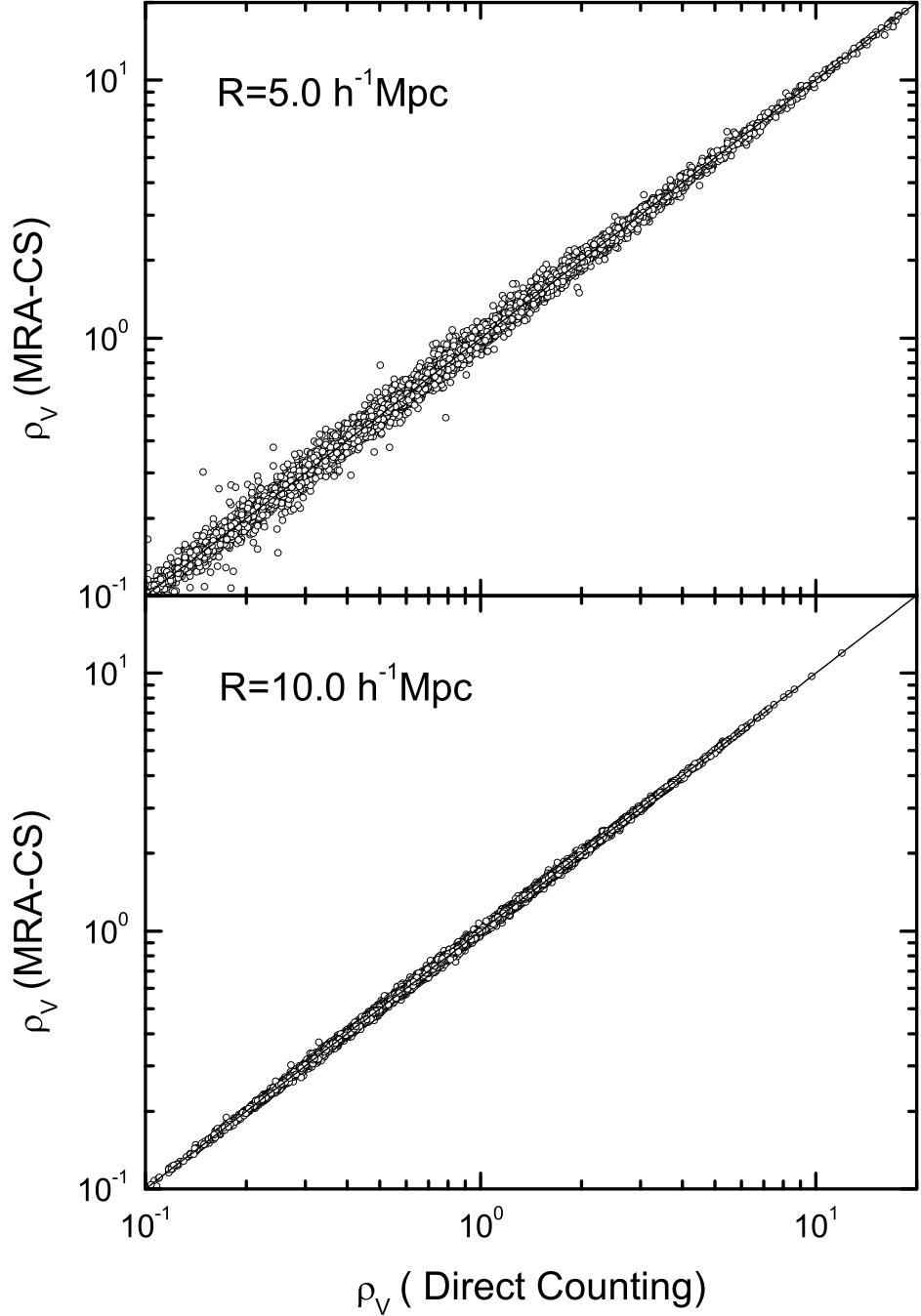}\vspace{-1.0cm}
\caption {Spherical counts in cells: the volume-averaged density
calculated by the MRA-CS scheme versus those from the direct
counting. The measurements were made in the Virgo simulation sample.
The $3 \times 10^4$ scatters (circule) randomly chosen from a total
$256^3$ sampling are displayed. The upper panel is for the cell size
$R=5$h$^{-1}$Mpc and the lower panel for $R=10$h$^{-1}$Mpc.}
%\end{center}
\end{figure}

%fig5
\begin{figure}
%\begin{center}
%\vspace{15.0cm}
%\special{psfile=f1.ps  hoffset=-0  voffset=-10 hscale=70 vscale=70}
\centering
\includegraphics[width=15.0cm,angle=0]{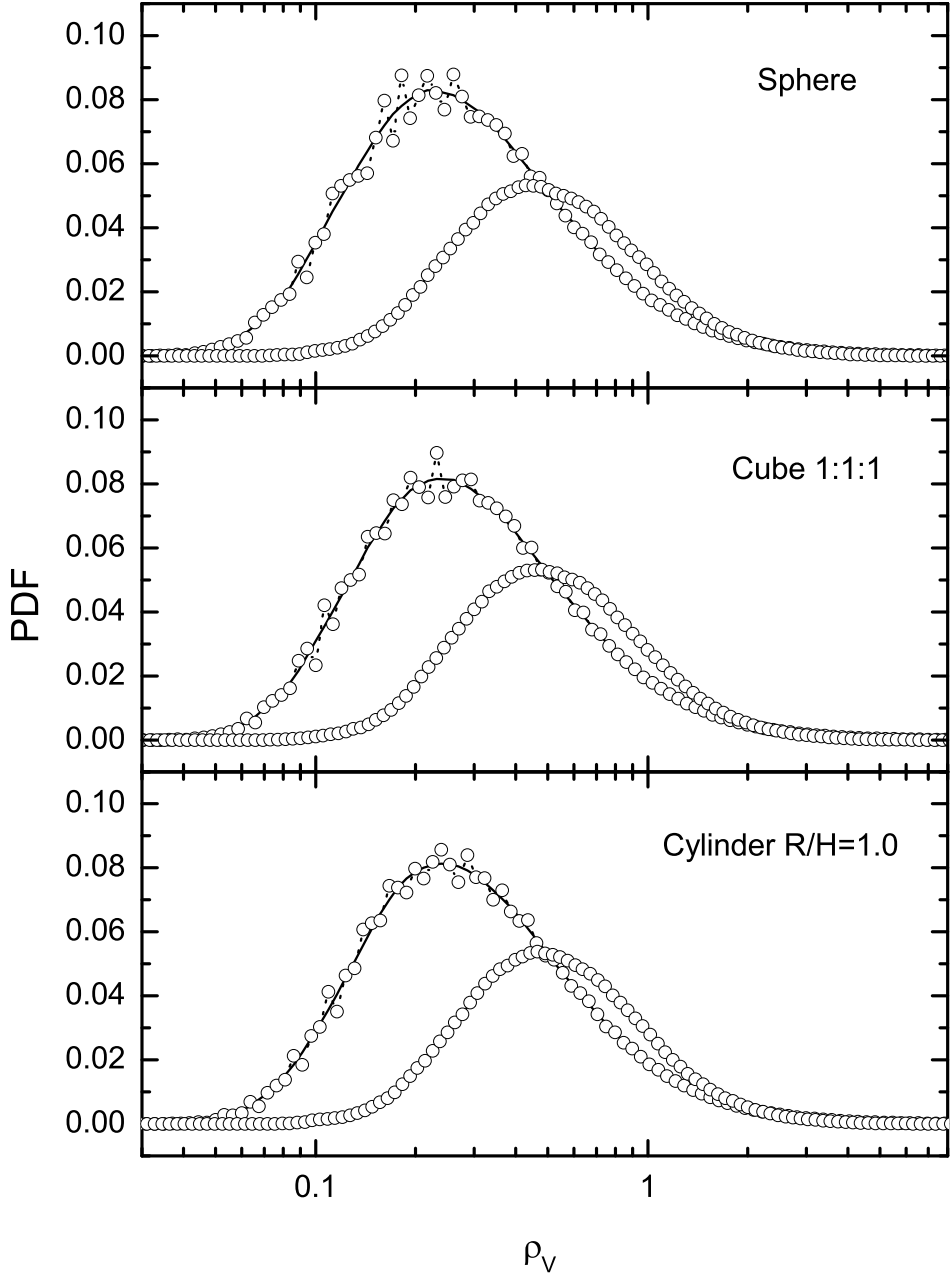}
\vspace{-1.0cm} \caption {Comparison of the probability distribution
function of volume-averaged density obtained from the MRA-CS
algorithm (solid line) with that by direct counting (circule). The
counts in cells are performed in the same simulation sample as in
fig.2. The upper, middle and lower panels display the results for
spherical, cubic and cylinder cells. In each panel, the right and
left curves denote results for cells with the equivalent radius
$R=5$h$^{-1}$Mpc and $R=10$h$^{-1}$Mpc respectively.}
%\end{center}
\end{figure}

%\fig6
\begin{figure}
%\begin{center}
%\vspace{16.0cm}
%\special{psfile=f1.ps  hoffset=-0  voffset=-10 hscale=70 vscale=70}
\centering
\includegraphics[width=18.0cm,angle=0]{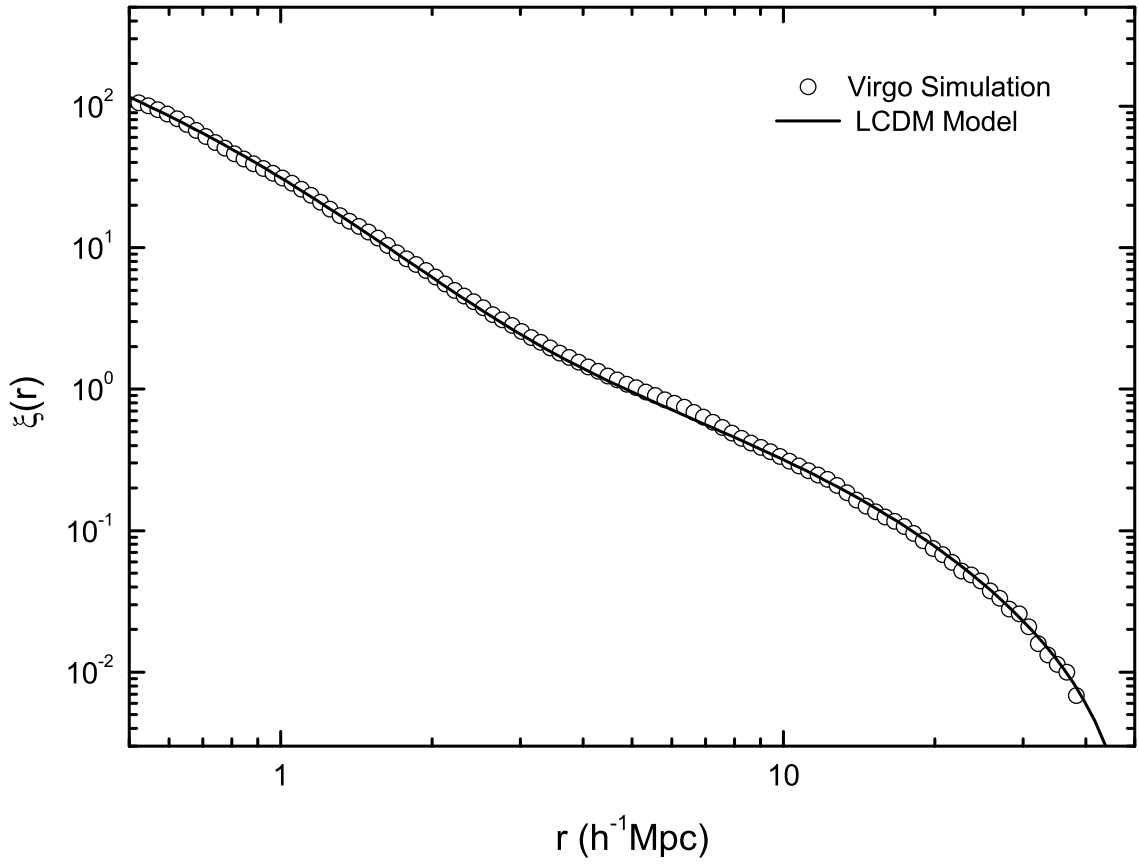}\vspace{-1.0cm}
\caption {The two-point correlations function measured with the
MRA-CS alogrithm (open circules) in the Virgo LCDM sample vs. the
nonlinear power spectrum given by Smith et.al (2003)(solid line). In
the MRA-CS calculation, a $512^3$ grid has been used.}
%\end{center}
\end{figure}

%\fig7
\begin{figure}
%\begin{center}
%\vspace{16.0cm}
%\special{psfile=f1.ps  hoffset=-0  voffset=-10 hscale=70 vscale=70}
\centering
\includegraphics[width=18.0cm,angle=0]{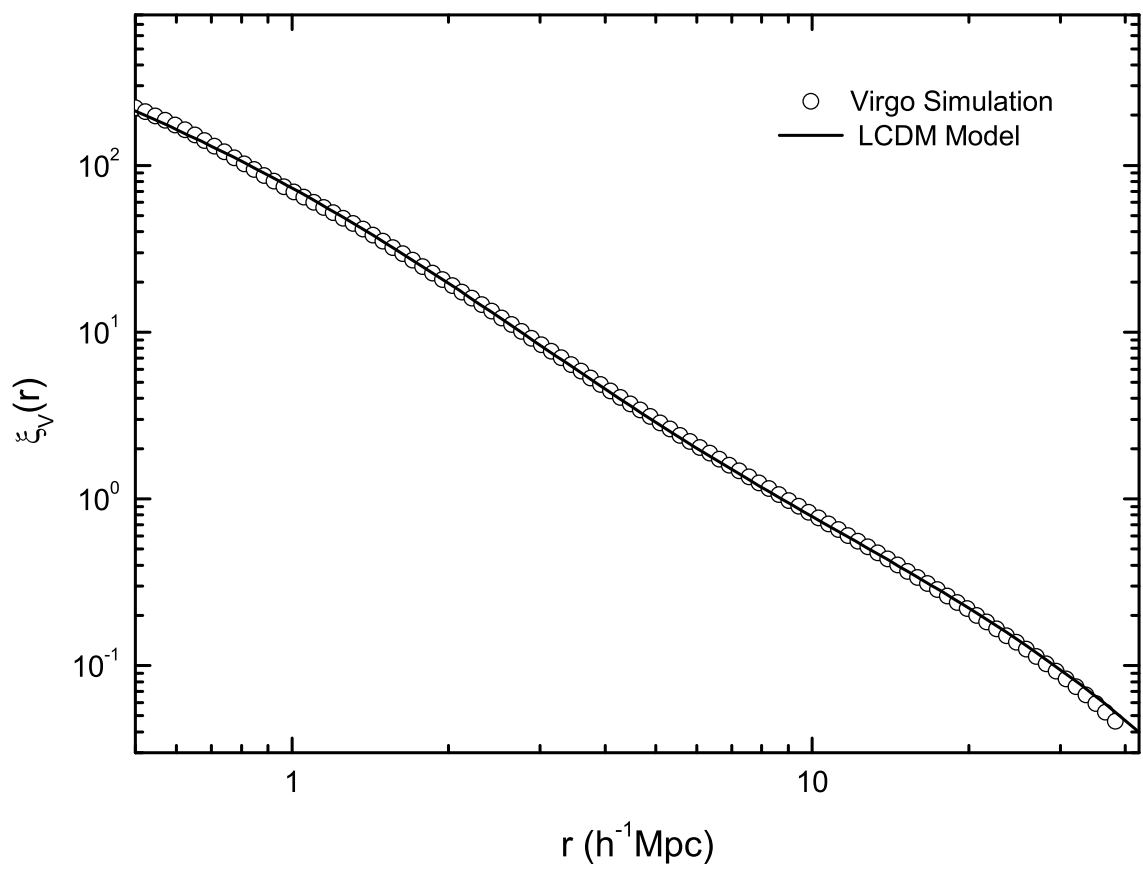}\vspace{-1.0cm}
\caption {Same as fig.6, but for the integral two-point correlation
function.}
%\end{center}
\end{figure}

%\fig8
\begin{figure}
%\begin{center}
%\vspace{16.0cm}
%\special{psfile=f1.ps  hoffset=-0  voffset=-10 hscale=70 vscale=70}
\centering
\includegraphics[width=18.0cm,angle=0]{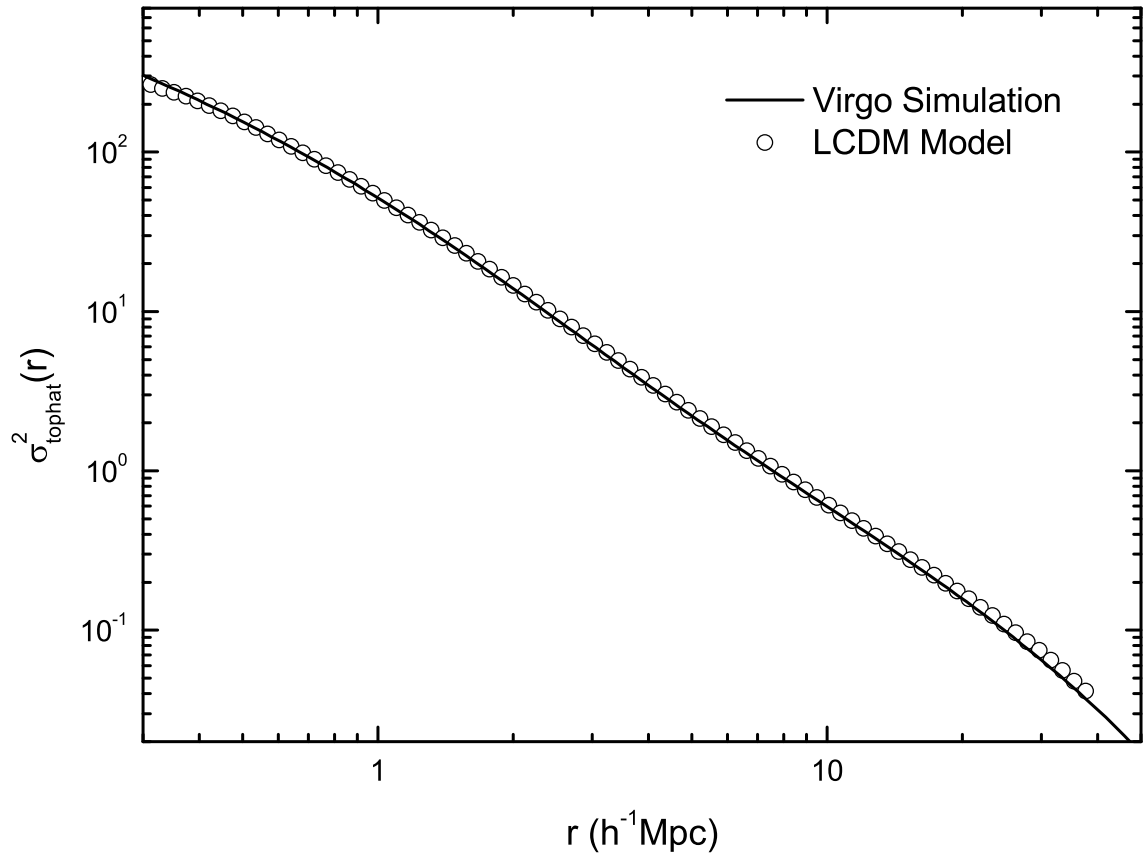}\vspace{-1.0cm}
\caption {Same as fig.6, but for the variance of the top-hat
filtered density field.}
%\end{center}
\end{figure}

\begin{figure}
%\begin{center}
%\vspace{16.0cm}
%\special{psfile=f1.ps  hoffset=-0  voffset=-10 hscale=70 vscale=70}
\centering
\includegraphics[width=18.0cm,angle=0]{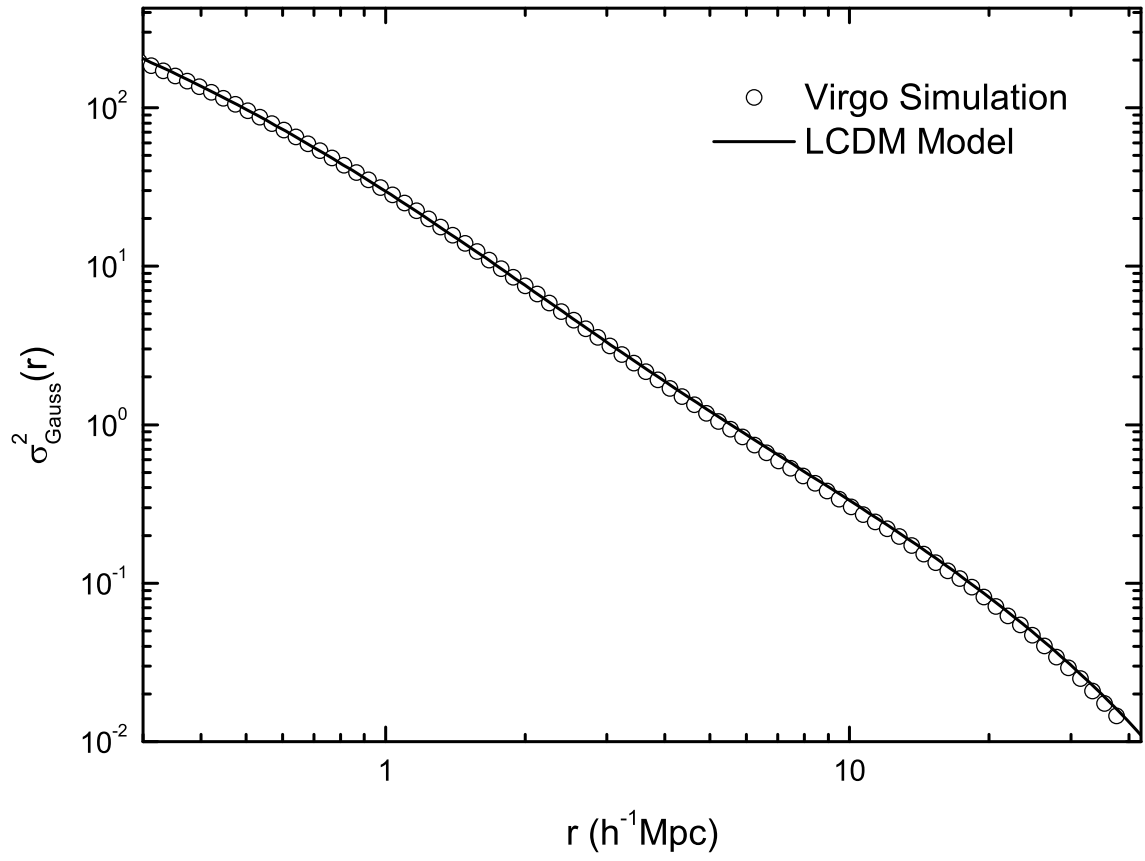}\vspace{-1.0cm}
\caption {Same as fig.6, but for the variance of the Gaussian
filtered density field.}
%\end{center}
\end{figure}

\end{document}